# A Retrospective Study on the Investigation of Potential Clinical Benefits of Online Adaptive Proton Therapy for Head and Neck Cancer

Chih-Wei Chang[1,#], Duncan Bohannon[1,#], Zhen Tian[2,*], Yinan Wang[1], Mark W. Mcdonald[1],

David S. Yu[1], Tian Liu[3], Jun Zhou[1,*] and Xiaofeng Yang[1,*]

[1]Department of Radiation Oncology and Winship Cancer Institute, Emory University, Atlanta, GA 30308

[2]Department of Radiation and Cellular Oncology, University of Chicago, Chicago, IL, 60637

[3]Department of Radiation Oncology, Mount Sinai Medical Center, New York, NY 10029

# Co-first author

*Co-corresponding author

**Email**: zhen.tian@uchicago.edu (ZT), jun.zhou@emory.edu (JZ) and xiaofeng.yang@emory.edu (XY)

## Abstract

**Purpose**: Proton therapy is very sensitive to patient anatomical changes that often occur in HN cancer patients during their treatment courses. Online adaptive proton therapy (APT) is an ideal solution theoretically, which however is challenging and not available in proton clinics yet. Although multiple groups have been endeavoring to develop online APT technology, there is a concern in the radiotherapy community about the necessity of online APT because of its unknown impact on treatment outcomes. Hence, we have performed a retrospective study to investigate the potential clinical effects of online APT for HN cancer patients in relative to the current offline APT via simulations, hoping to help the community to evaluate the need of realizing online APT.

**Methods**: Our retrospective study was conducted using the clinical data of four HN cancer patients, who were treated with proton therapy at our institution and had at least one offline plan adaptation during their treatment courses. To mimic an online APT treatment course, we have recalculated and evaluated the actual dose of the current treatment plan on patient's actual treatment anatomy captured by cone beam CT (CBCT) for each fraction, and adapted the plan when warranted, assuming that these adaptions were performed online before treatment delivery. The cumulative dose of this simulated online APT course was compared with that of the actual offline APT course, as well as with the initially designed plan dose of the original treatment plan generated prior to the treatment course. Tumor control probability (TCP) and normal tissue complication probability (NTCP) were also calculated to estimate the potential effect of online APT on treatment outcome.

**Results**: For patients 1 and 2, with clinically comparable sparing of organs at risk (OARs), the simulated online ART course maintained a relatively higher CTV dose coverages than the offline ART course, particularly for CTV-Low, which led to an improvement of 2.66% and 4.52% in TCP of CTV-Low. For patients 3 and 4, with clinically comparable CTV dose coverages, the simulated online ART course achieved better OAR sparing than the offline ART course. The mean doses of right parotid and oral cavity were decreased from 29.52 Gy relative biological effectiveness (RBE) and 41.89 Gy RBE to 22.16 Gy RBE and 34.61 Gy RBE, respectively, for patient 3, leading to a reduce of 1.67% and 3.40% in NTCP for the two organs. The mean dose of right parotid was decreased from 21.71 Gy RBE to 19.37 Gy RBE for patient 4, leading to a reduce of 0.73% in NTCP for right parotid.

**Conclusion**: Compared to the current clinical practice, the retrospective study indicated that online APT tended to spare more normal tissues by achieving the clinical goal with merely half of the positional uncertainty margin. With the results of this study, we expect that many HN cancer patients will benefit from online APT, receiving higher and homogenous target coverage and/or better sparing of critical organs, which might lead to improved local tumor control and reduced treatment toxicity. Future studies are needed to help identify the patients with large potential benefits prior to treatment in order to conserve scarce clinical resources.

## 1. Introduction

Head and neck (HN) cancer is a heterogeneous group of malignancies and accounts for about 4% of all cancers in the United States, with an annual estimation of 66,470 cases and 15,050 deaths [1]. The prevailing complication for standard radiation treatment of HN cancer is the intrinsic promixity of multiple critical organs that are highly radiosensitive (e.g., brain, brainstem, cranial nerves, spine, salivary glands) and hence the potential significant detrimental effects on functionality (e.g., the loss of sensory function, imparied salivary production and swallowing capacity, and neurological damage) and the risk of secondary cancers. Enormous efforts have been made to revolutionize photon radiotherapy (RT) in the past two decades, such as the invention of intensity modulated RT (IMRT) and volumetric modulated arc therapy (VMAT) [2-6], resulting in highly conformal radiation dose to the target volumes with much improved sparing of adjacent critical organs. Nevertheless, the inherent physics properties of photon RT make it often unavoidable to expose the critical organs with low to moderate doses in HN cancer treatment in order to adequately treat the malignancy. Unlike the photon RT, the Bragg peak effect is one of the important characteristics of proton RT, offering unique advantages in critical organ sparing for HN cancer treatment. It has been reported that compared with IMRT and VMAT, intensity modulated proton therapy (IMPT) can significantly reduces the maximum doses to the spinal cord and brainstem and the mean doses to the larynx and parotid glands [7-9]. These dosimetric benefits of IMPT have been observed to translate to treatment outcome benefits of reduced toxicity and improved quality of life in HN cancer patients [8, 10-15].

While, the potential benefits of proton RT for HN cancer are not yet fully realized. The treatment plans are typically designed based on a computed tomography (CT) scan acquired a few weeks before treatment and the treatment is delivered over 4-6 weeks, the HN anatomy is prone to considerable volumetric and positional change over this timeframe, due to resolving postoperative edema, tumor shrinkage, changes in overall body weight, and daily treatment setup variations. For instance, the gross tumor volume (GTV) decreases at a median rate of 1.8% per day, which is frequently asymmetric and hence also results in a displacement of the center of tumor mass [16]. The regression of HN tumors also leads to the shrinkage and shift of parotid glands and submandibular. Proton RT is very senstive to the anatomical changes, as the range of Bragg peak highly depends on the tissue density along the beam path [17-20]. The original plan based on the patient's previous anatomy can result in mispositioned Bragg peaks in patient's current anatomy, leading to underdosing of tumors and/or overdosing of the critical organs.

Although robust plan optimization which considers multiple possible treatment scenarios during the plan optimization [21, 22] can account for the anatomical changes to some extent, this method can maintain the designed plan quality only when the patient's actual anatomy in treatment is covered by the worst scenarios during robust optimization [23-25]. Clinical studies have found that robust optimization is not sufficient for many HN cases [26-28]. Besides, apart from the elongated planning time due to the much more intensive computational tasks to consider each scenario in the plan optimization [29], robust plan optimization usually softens dose gradients, resulting in enlarged high dose volumes and higher dose to critical organs and normal tissues compared to the plans optimized with the norminal scenario alone [30, 31]. Adaptive proton therapy (APT) offers a general solution to account for patient anatomical changes without increasing dose to critical organs and normal tissues [32]. Many proton centers have started to acquire weekly quality assurance CT (QACT) images to regularly monitor the anatomical changes of HN cancer patients during their treatment courses and, if needed, perform offline replanning based on the QACT images. It has been reported that compared to non-adaptive IMPT, offline adaptive IMPT significantly reduces the maximum doses to the spinal cord and mandible and the mean doses to the larynx and ipsilateral parotid gland [7]. However, due to the intensive work required by the current APT technology, it typically takes about a week from the acquisition of QACT to the time when the new treatment plan is ready for review. Meanwhile, although failing the original clinical goals on patient's current anatomy, the old plan often has to be continually used until the new plan is ready because of the danger of accelerated tumor repopulation caused by breaks in the treatment course.

The ideal and desired solution is online APT, in which the actual plan quality of the original plan on patient's up-to-date anatomy after treatment setup is assessed at each fraction and, when needed, rapid replanning is performed for the current fraction while the patient is lying on the treatment couch [33]. However, due to the lack of high-quality on-board 3D imaging system and the long time required for a complex and labor-intensive workflow [34], online APT is challenging and still not available in proton clinics yet, in contrast to the successful implementation of online adaptive photon therapy such as MR-Linacs and the ring-shape CBCT-Linac Halcyon® along with Ethos™ treatment planning system (Varian Medical Systems, Palo Alto, CA) as well as the recent boom of their adoption in photon clinics. Although multiple groups have been endeavoring to overcome these challenges to enable online APT [35-42], there is also a concern in the RT community about the necessity of online APT because of its unknown impact on treatment outcomes. Theoretically speaking, the online APT, if successfully developed and clinically implemented in near future, will guarantee the optimal plan quality at every treatment fraction, dispite the anatomical changes and daily treatment setup variations. However, in the real world, it is possible that the dosimetric disadvantages of the current offline APT shown in ceartin treatment fractions might be smeared due to the daily setup variations and hence counteracted in the cumulative dose over the entire treatment course, which might dampen the need of developing the online APT technology.

Thereby, the purpose of this study is trying to investigate the potential clinical impact of online APT for patients with anatomical changes during their RT courses and compared with the current offline APT practice. As the online APT technology is currently not available, we have performed a retrospective study on four HN cancer patients, who were treated with proton therapy at our institution and had at least one offline plan adaptation during their treatment courses. To mimic an online APT treatment course, in this study we have recaulculated and evaluated the actual dose of the current in-use treatment plan (i.e., the original treatment plan generated prior to the treatment course if no plan adaptation hasn't been triggered yet or the latest adapted plan) on patient's actual anatomy captured by cone beam CT (CBCT) at each fraction, and when warranted, adapted the current plan based on patient's up-to-date anatomy, assuming that these adaptions were performed online before treatment delivery. The cumulative dose of this simulated "online" APT course was compared with the cumulative dose of the offline APT course that was used for the actual patient treatment, as well as with the initially designed plan dose of the original treatment plan generated prior to the treatment course (referred to as the nominal plan). Tumor control probability (TCP) and normal tissue complication probability (NTCP) were also calculated and compared for the two APT courses to estimate the potential effect of online APT on treatment outcome. We hope that this simulation study might shed some lights on the potentials of online APT to better help the RT community and vendors determine whether it is worth devoting great efforts to realize online APT.

## 2. Materials and Methods
### 2.1 Patient data and the current offline APT workflow

We have retrieved clinical data of four HN cancer patients, who were treated with proton therapy at our institution and had at least one plan adaptation during their treatment courses, from our institutional database for the retrospective study. In their actual offline APT treatment, the CT images that were acquired prior to the start of the treatment course and used for the nominal treatment planning (referred to as planning CT) were acquired using a Siemens SOMATOM Definition Edge scanner. The treatment planning system (TPS), RayStation 9A (RaySearch Lab., Stockholm, Sweden), was used to perform the nominal treatment planning following our institutional standard protocol. The RayStation is running on a clinical server with dual Intel® Xeon® Gold 6136 CPU (2017 model year), 512GB RAM, and an NVIDIA Quadro RTX 8000. It supports GPU-based Monte Carlo dose calculation, and deformable image registration. All clinical plans were robustly optimized with 3.0 mm setup positional uncertainty in the orthogonal directions and 3.5% range uncertainty, resulting in a total of 21 scenarios.

In our institution, most treatment plans for HN proton RT were optimized using five proton beams, including left posterior oblique (LPO), left anterior oblique (LAP), anteroposterior (AP), right anterior oblique (RAO), and right posterior oblique (RPO). A constant relative biological effectiveness (RBE) of 1.1 was assumed for proton radiation doses based on IAEA/ICRU guidelines [43, 44]. The prescribed doses ranged from 68.4-70 Gy (RBE) with 1.9-2.0 Gy (RBE) daily fractions. During the treatment course, bi-weekly quality assurance CT (QACT) images were obtained from the identical CT scanner to monitor the patient's anatomical change and assess the plan quality of the nominal plan on the new anatomy captured by the QACT. Offline replanning was triggered if the obtained dose volume metrics of the recalculated nominal plan didn't satisfy our institutional requirement any longer. For the offline replanning, a deformable registration between the planning CT and the recent QACT was first performed to propagate the contours of the clinical target volumes (CTVs) and the organs at risk (OARs) from the planning CT to the QACT. After the attending physician reviewed the propagated contours and modified the contours when needed, replanning was then performed by one of our proton dosimetrists using RayStation following the same planning protocol used in the nominal (initial and original) treatment planning. At the beginning of each treatment fraction, CBCT images were acquired to guide the treatment setup. Table 1 summarizes the treatment information of the actual offline APT treatment for the four HN patients.

**Table 1.** Summary of the actual treatment information of the four HN patients.

|  |  | Patient 1 | Patient 2 | Patient 3 | Patient 4 |
|---|---|---|---|---|---|
| High-Risk Clinical target volume (CTV-High) | Initial volume | 60.36 cm3 | 84.40 cm3 | 68.82 cm3 | 57.67 cm3 |
|  | Final volume | 61.87 cm$^3$ | 78.73 cm$^3$ | 58.46 cm$^3$ | 59.04 cm$^3$ |
| Intermediate-Risk Clinical target volume (CTV-Mid) | Initial volume | 145.59 cm3 | 122.69 cm3 | 132.89 cm3 | 84.46 cm3 |
|  | Final volume | 144.37 cm$^3$ | 115.39 cm$^3$ | 136.69 cm$^3$ | 81.94 cm$^3$ |
| Low-Risk Clinical target volume (CTV-Low) | Initial volume | 404.53 cm3 | 219.41 cm3 | 180.18 cm3 | 270.44 cm3 |
|  | Final volume | 405.59 cm$^3$ | 189.75 cm$^3$ | 160.75 cm$^3$ | 243.47 cm$^3$ |
| Prescription dose (Gy-RBE) | CTV-High | 68.4 | 70.0 | 70.0 | 70.0 |
|  | CTV-Mid | 64.8 | 63.0 | 60.2 | 63.0 |
|  | CTV-Low | 54.0 | 56.0 | 53.9 | 56.0 |
| Number of fractions |  | 36 | 36 | 35 | 35 |
| Replan times |  | 2 | 2 | 1 | 1 |

### 2.2 Proposed workflow of CBCT-guided online APT

Figure 1 depicts that our proposed workflow of the CBCT-guided online APT for the existing proton systems, which consists of the following steps.

Step 1: At the beginning of each treatment fraction, the CBCT images will be acquired and registered to the planning CT images via rigid image registration to determine the couch movement that is needed to reproduce the planned treatment position, which is same as the current offline APT.

Step 2: The CBCT images and the subsequent couch movement values will then be retrieved from the ARIA® oncology information system (V16.1, Varian Co., Palo Alto, CA) through an internal network and imported into RayStation. Note that the couch movement will be applied on the retrieved CBCT images to ensure that the actual treatment setup with patient's actual anatomy will be used to evaluate the plan quality of the current in-use plan.

Step 3: The acquired CBCT images cannot be directly used for daily dose evaluation and online plan adaptation, as the existing on-board CBCT imaging system suffers from severe image artifacts mainly due to scatter contamination, which impair the soft-tissue contrast and lead to large Hounsfield Unit (HU) uncertainties. Hence, we will perform CBCT image correction at this step. Specifically, we will deform the planning CT images to the daily CBCT images via deformable image registration in order to use the correct

HU numbers from the planning CT images while preserving the patient's actual anatomy captured on the CBCT images. Please note that by doing this, the region outside the CBCT field of view (FOV) will also be extended by the planning CT images for daily dose evaluation. In addition, as the nasal cavity filling can vary from time to time, we will copy the air cavities from the CBCT images to the deformed planning CT images to preserve the actual nasal cavity filling. The resulting final images are referred to as corrected CBCT images in the rest of this manuscript.

Step 4: We will propagate the contours of the CTVs and OARs from the planning CT images to the corrected CBCT images using RayStation, similar to the contour propagation from the planning CT images to the QACT images in the current offline APT practice. Note that users have the flexibility to specify the propagation of each contour to be either a rigid copy or a deformable mapping in RayStation.

Step 5: Attending physician will review the propagated contours and modify the contours if needed (the final contours are referred to as daily contours in the rest of this manuscript).

Step 6: We will recalculate the dose of the current in-use treatment plan on the corrected daily CBCT images using Monte Carlo dose calculation.

Step 7: We will calculate the dose volume histograms (DVHs) for each CTV and OAR using the daily contours, and evaluate whether the plan quality will still satisfy our clinical goals. If not, online adaptation will be triggered to conduct steps 8-11.

Step 8: If online adaptation is triggered, a treatment plan optimization will be performed using the corrected CBCT images and the daily contours to generate a new plan that will satisfy the clinical goals.

Step 9: Attending physician will review the plan and modify the plan if needed.

Step 10: The approved new plan will be exported to the treatment delivery system to generate a pre-delivery machine log file with sequences of the planned machine parameters. Secondary dose calculation will be performed as a patient-specific quality assurance for online adaptation, using this pre-delivery log file and the corrected CBCT images, to ensure patient safety.

Step 11: Another CBCT scan will be acquired before treatment delivery of the new plan to ensure the same treatment positioning as on the corrected CBCT images, in case that the patient might move while waiting for the new plan.

Step 12: Treatment delivery.

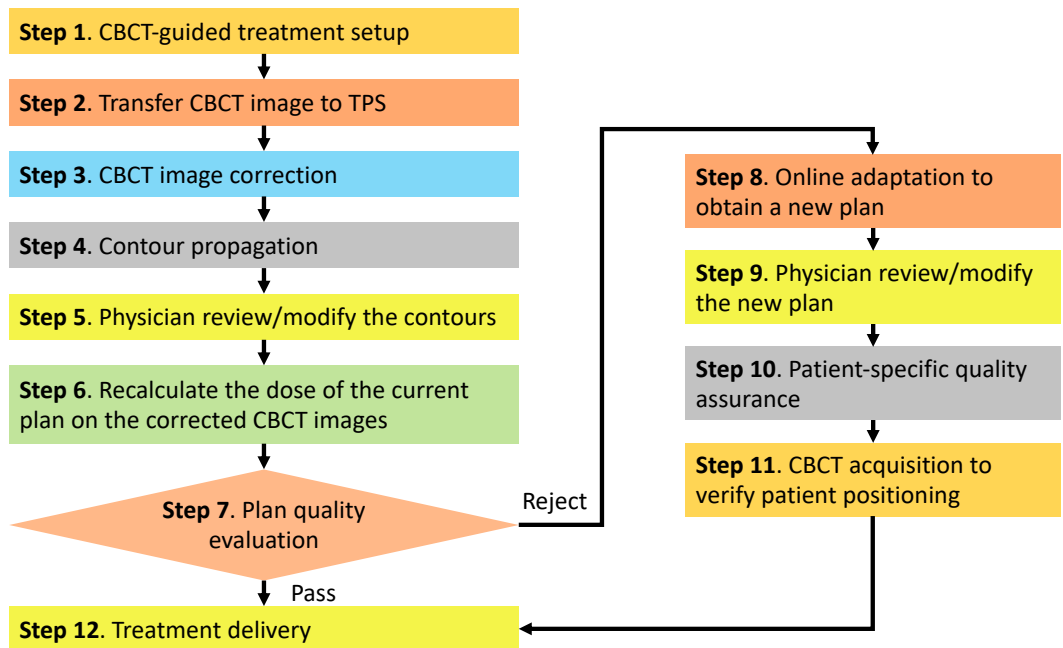

**Figure 1**. The proposed workflow of CBCT-guided online APT used in this retrospective study.

**2.3 Retrospective study**

As mentioned in the introduction session, our purpose of this retrospective study is trying to investigate the potential clinical effect of online APT, if successfully developed in future, for HN cancer patients with observed anatomical changes during the RT course, compared with the current offline APT practice. Because online APT is not available in proton clinics yet, in our study we have imported the daily CBCT images, that were acquired from each treatment fraction of the actual offline APT treatment of the four HN patients, into RayStation to conduct steps 3-10 of the proposed workflow to mimic an online APT treatment course. In-house Python scripts have been developed to automatically streamline the workflow in the TPS. When the online plan adaptation was triggered in our simulation, one of our experienced proton dosimetrists performed a replanning using the corresponding corrected CBCT images and the daily contours. In our retrospective study, this replanning process was regarded to be a "online" replanning process, assuming that automatic and rapid online replanning will be realized via advanced artificial intelligence technologies in future to achieve sufficient plan quality that is comparable to human planners.

Specifically, the same beam arrangement that was used in the nominal treatment planning of the same patient was used in our replanning. The dosimetrist used the same dose constraints that were used for the nominal treatment planning as a reasonable starting point of replanning, and adjusted the dose constraints repeatedly, as we did to generate the nominal plan, to achieve a satisfying new plan. Unlike the nominal treatment planning or the offline replanning, we didn't use 21 scenarios in the 3D robust optimization for the replanning in our study, as the actual treatment setup at that specific treatment fraction was already known and present in the corrected CBCT images. Instead, only range uncertainties (3 scenarios, i.e., nominal and

±3.5% range uncertainty) were included in the robust optimization, which significantly reduced the replanning time as well. In addition, a 1.5 mm margin was added to the CTV volumes for plan optimization to account for the potential intrafraction motion of HN cancer patients, which has been found to be within 1 mm translation and 0.7 degree rotation with the aid of immobilization devices [45], as well as the spot position uncertainty of scanning proton pencil beam, which has been found to be within 1 mm for the maximum day-to-day variance for any given spot positions [46].

After simulating the CBCT-guided online APT for every treatment fraction by conducting the steps 3-10 of the proposed workflow, deformable image registration was performed to deform the corrected CBCT images to the planning CT images to accumulate the actual dose of the current in-use plan (i.e., either the original plan if online adaptation was not triggered yet or the latest new plan) over the entire treatment course. For comparison purposes, the actual dose of the plan, that was delivered in each treatment fraction of the actual offline APT treatment course, was recalculated in the corresponding corrected CBCT images and accumulated over the entire offline APT course. In our study, we have compared these two cumulative dose distributions of the virtual "online" APT treatment course and the actual offline APT treatment course for the four HN cases, with respective to the originally planned dose of the nominal treatment plan, to investigate the potential dosimetric effect of online APT.

**2.4 TCP and NTCP model**

To investigate the potential effect of online APT on treatment outcome, we implemented an equivalent uniform dose (EUD)-based model [47] to estimate TCP and NTCP for both treatment courses. Table 2 gives the essential parameters for the EUD-based TCP/NTCP model. All tumor and normal tissue parameters were acquired from literature [48-54].

**Table 2.** Tumor and normal tissue parameters for EUD-based TCP and NTCP estimation.

|  | a | $\gamma_{50}$ | $TCD_{50}/TD_{50}$ (Gy) | $\alpha/\beta$ (Gy) |
|---|---|---|---|---|
| Tumor | -8 | 2 | 50 | 10 |
| Parotid | 5 | 4 | 46 | 3 |
| Oral cavity | 5 | 4 | 46 | 3 |

**3. Results**

**3.1. Comparisons of online and offline adpative treatment planning**

Figures 2-5 show the cumulative doses of the simulated online APT course and the actual offline APT course, as well as the designed dose of the nominal plan for the four patients. The nominal plan serves as a baseline to evaluate the actual efficacy of the online and offline APT courses. To quantitatively compare these three dose distributions, multiple dosimetric endpoints of CTVs and the involved OARs, that are used in our institution for plan quality evaluation, were calculated and listed in Table 3 for the four patients.

For patient 1, it can be observed from Figure 2(c) that the dose coverages of all the three CTVs got deterioated in the actual offline APT course, with some hot spots (i.e., high dose regions) appeared inside the CTV-High. The DVHs shown in Figure 2(d) also indicate that the offline APT course had worse dose coverages for all the three CTVs, compared to the nominal plan. Besides, although the dose to the right parotid was reduced in the offline APT course, the dose to the left parotid got increased. In constrast, the online APT course achieves comparable dose coverages for CTV-Mid and CTV-Low and acceptable coverage for CTV-High, while slightly improving the sparing of both left and right parotids and oral cavity, compared to the nominal plan. This is also reflected in the obtained dose volume endpoints listed in Table 3. Specifically, the CTV dose coverages (i.e., $V_{100}$ which denotes the percentage CTV volume that is covered by the prescription dose) achieved by online APT were 93.29%, 96.50%, 99.32% for CTV-High, CTV-Mid and CTV-Low, respectively, compared to 92.07%, 94.72%, 97.88% obtained by offline APT and 94.8%, 96.25%, 99.12% that were originally designed in the nominal plan. The mean doses of the right parotid, left parotid and oral cavity obtained by online APT were 2.07 Gy (RBE), 18.49 Gy (RBE), and 6.44 Gy (RBE), compared to 1.64 Gy (RBE), 19.60 Gy (RBE), and 6.53 Gy (RBE) obtained by offline APT and 2.36 Gy (RBE), 18.86 Gy (RBE) and 6.55 Gy (RBE) designed in the nominal plan.

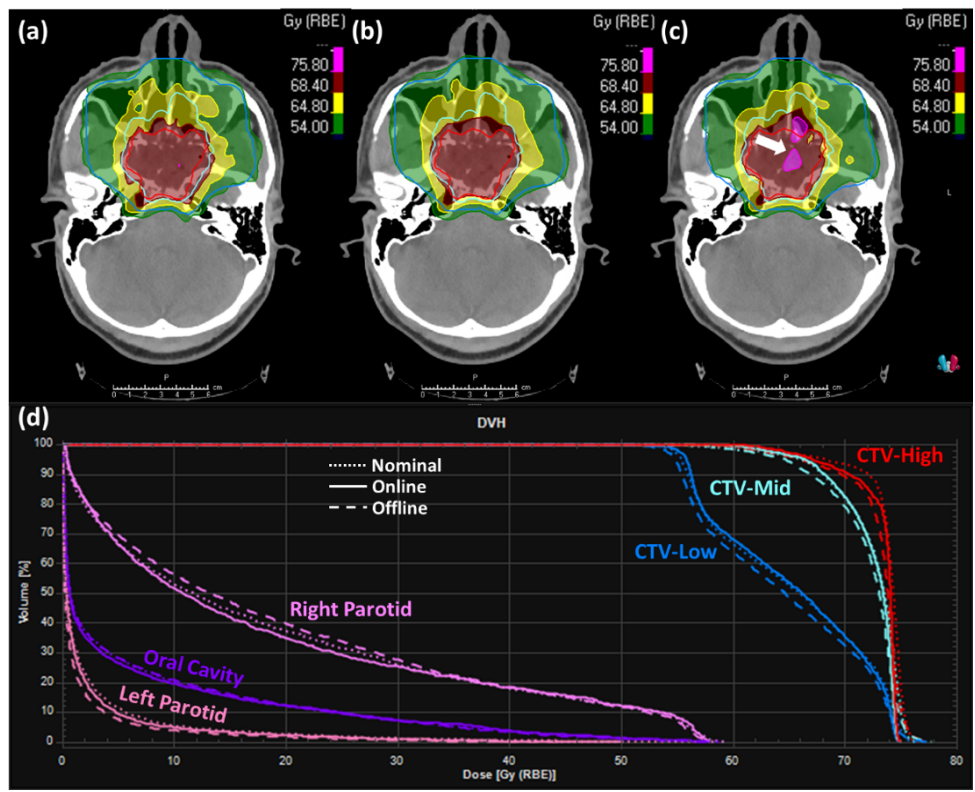

**Figure 2**. Dose comparison for patient 1 among the cumulative doses of the simulated online APT course, the actual offline APT course, and the originally designed dose of the nominal plan. The top row displays the dose distributions in color wash for the nominal plan (a), the online APT course (b), and the offline APT course (c), overlaid with transverse CT images and the contours of CTVs and OARs shown in lines. The bottom row (d) displays the Dose-volume histograms (DVHs) of CTVs and involved OARs, where the dot, solid, and dash lines represent the nominal plan, the online and offline APT courses, respectively. The white arrow indicates the hot spot location.

For patient 2, Figure 3 have clearly shown that with comparable OAR sparing, the actual dose coverages obtained by offline APT got deteriorated for all the three CTVs, particularly for CTV-Low whose coverage ($V_{100}$) became 89.53%, compared to 99.08% that was designed in the nominal plan. Meanwhile, the offline APT resulted in more hot spots of higher doses inside CTV-High. In contrast, the simulated online APT course achieved slightly better dose coverages for CTV-High (i.e., 98.88% vs. 98.20% in the nominal plan) and CTV-Mid (i.e., 99.81% vs. 99.17% in the nominal plan) and comparable dose coverage for CTV-Low (i.e., 98.97% vs. 99.08% in the nominal plan), and meanwhile slightly reduced the mean doses of right parotid, left parotid, and oral cavity (i.e., 15.66, 23.50 and 29.34 Gy (RBE)) vs. 18.80, 24.62 and 30.30 Gy (RBE)).

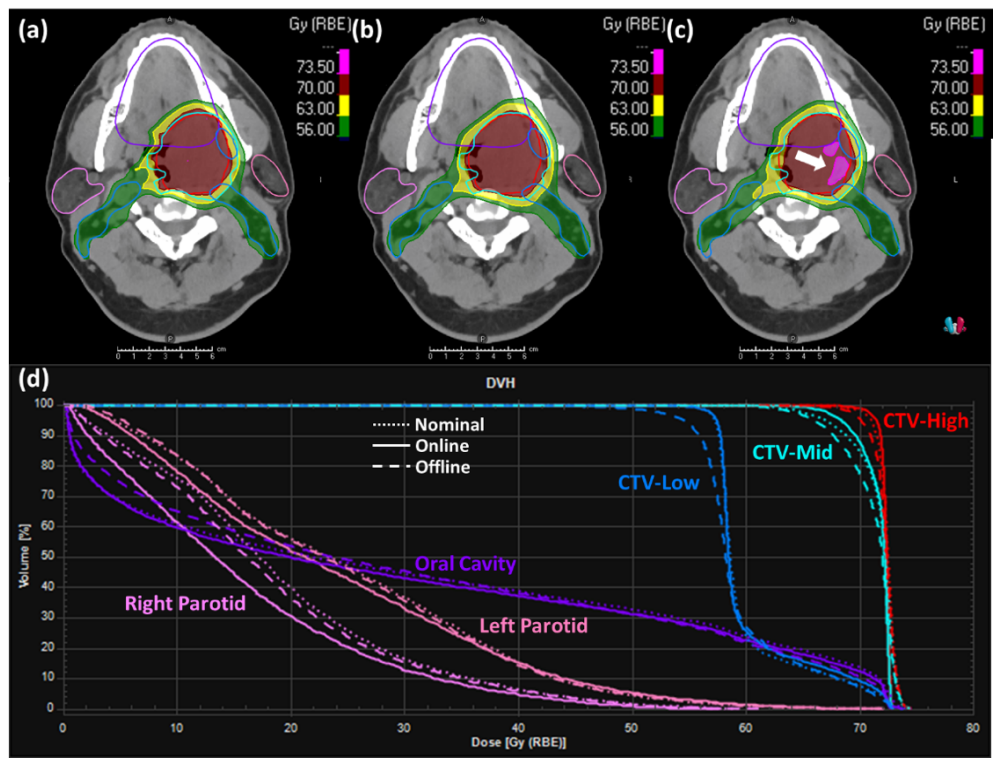

**Figure 3**. Dose comparison for patient 2 among the cumulative doses of the simulated online APT course, the actual offline APT course, and the originally designed dose of the nominal plan. The top row displays the dose distributions in color wash for the nominal plan (a), the online APT course (b), and the offline APT course (c), overlaid with transverse CT images and the contours of CTVs and OARs shown in lines. The bottom row (d) displays the Dose-volume histograms (DVHs) of CTVs and involved OARs, where the dot, solid, and dash lines represent the nominal plan, the online and offline APT courses, respectively. The white arrow indicates the hot spot location.

For patient 3, although maintaining the dose coverage of CTV-Low very comparable to the nominal plan, the offline APT course got worse coverages for the other two CTVs, particularly for CTV-Mid (i.e., 93.83% vs. 98.47% in the nominal plan). Meanwhile, the mean doses of right parotid, left parotid and oral cavity got increased from 28.58, 19.17 and 37.12 Gy (RBE) to 29.52, 20.12, and 41.89 Gy (RBE), respectively. For this patient, when compared to the nominal plan, the simulated online APT course got comparable dose coverage for CTV-Low and slightly worse dose coverages for CTV-High and CTV-Mid (i.e., 98.82% and 97.62% vs. 99.89% and 98.47% in the nominal plan), while significantly reducing the mean dose of right parotid, left parotid and oral cavity from 28.58, 19.17 and 37.12 Gy (RBE) to 22.16, 15.87 and 34.61 Gy (RBE).

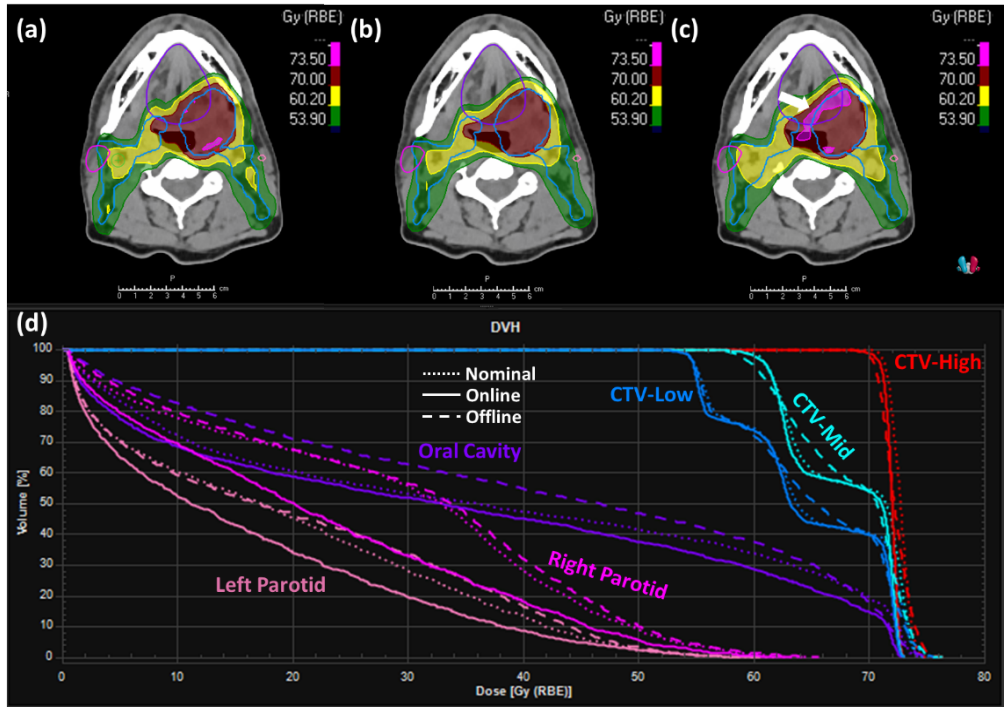

**Figure 4** Dose comparison for patient 3 among the cumulative doses of the simulated online APT course, the actual offline APT course, and the originally designed dose of the nominal plan. The top row displays the dose distributions in color wash for the nominal plan (a), the online APT course (b), and the offline APT course (c), overlaid with transverse CT images and the contours of CTVs and OARs shown in lines. The bottom row (d) displays the Dose-volume histograms (DVHs) of CTVs and involved OARs, where the dot, solid, and dash lines represent the nominal plan, the online and offline APT courses, respectively.

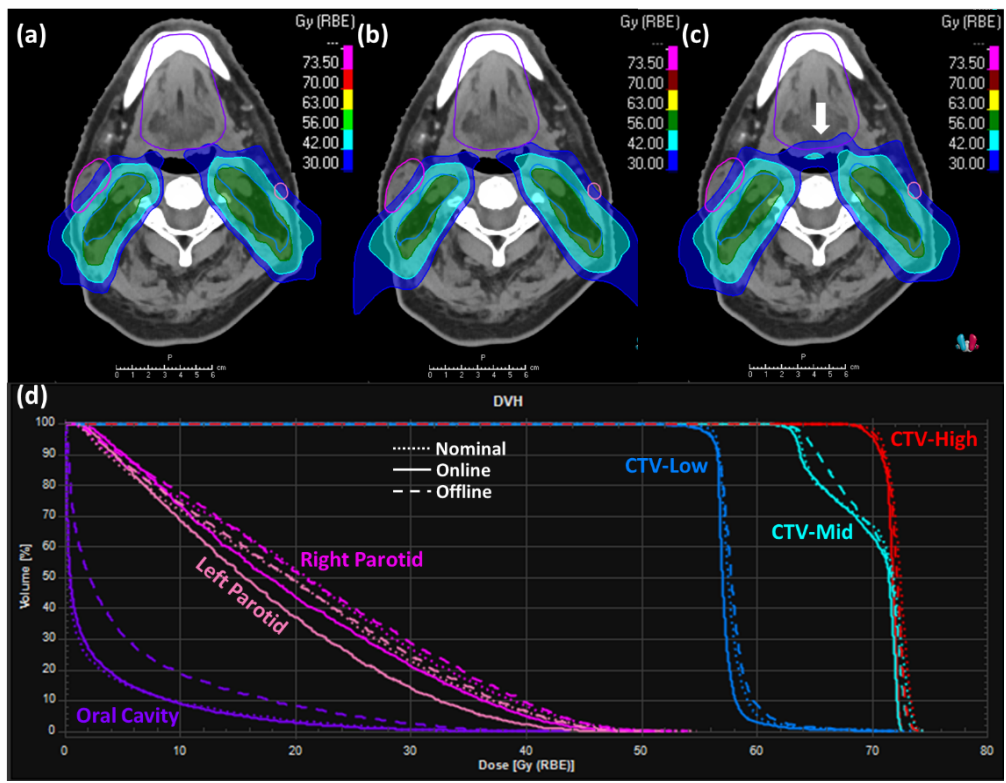

**Figure 5**. Dose comparison for patient 4 among the cumulative doses of the simulated online APT course, the actual offline APT course, and the originally designed dose of the nominal plan. The top row displays the dose distributions in color wash for the nominal plan (a), the online APT course (b), and the offline APT course (c), overlaid with transverse CT images and the contours of CTVs and OARs shown in lines. The bottom row (d) displays the Dose-volume histograms (DVHs) of CTVs and involved OARs, where the dot, solid, and dash lines represent the nominal plan, the online and offline APT courses, respectively. The white arrow indicates the dose escalation within the oral cavity from offline APT planning.

For patient 4, compared to the nominal plan, the offline APT obtained comparable dose coverage for CTV-Mid and lower coverage for CTV-High and CTV-Low, slightly higher mean dose for both parotids, and much higher mean dose for oral cavity (i.e., 5.72 Gy RBE vs. 2.83 Gy RBE). The online APT also obtained lower coverage for the CTVs compared to the nominal plan, it reduced the mean dose of the right parotid and left parotid from 21.19 and 19.81 Gy (RBE) to 19.37 and 17.15 Gy (RBE), respectively. With different trade-offs between CTV dose coverage and OAR sparing shown in the three dose groups, it is difficult to determine whether online APT presents a better treatment quality than the offline APT using the dosimetric endpoints only.

**Table 3.** Comparisons of dosimetric endpoints of CTVs and OARs among the nominal plan, the cumulative dose of the online APT course and the cumulative dose of the offline APT course for the four patients. Here, $V_{100}$ denotes the CTV percentage volume that received 100% of the prescription does, $D_{98}$ and $D_{95}$ denote the minimum doses received by at least 98% CTV volume and 95% CTV volume, respectively. $D_{mean}$ denote the mean dose received by an OAR. $D_{max}/D_{RX}$ denotes the ratio of the maximum dose to the prescription dose of CTV-High ($D_{RX}$). Please note that the unit of all the doses reported in this table and in this manuscript are Gy (RBE).

| Patient 1 | CTV-High | | | CTV-Mid | | | CTV-Low | | | Right Parotid | Left Parotid | Oral Cavity | Body |
|---|---|---|---|---|---|---|---|---|---|---|---|---|---|
| | $V_{100}$(%) | $D_{98}$(Gy) | $D_{95}$(Gy) | $V_{100}$(%) | $D_{98}$(Gy) | $D_{95}$(Gy) | $V_{100}$(%) | $D_{98}$(Gy) | $D_{95}$(Gy) | $D_{mean}$(Gy) | $D_{mean}$(Gy) | $D_{mean}$(Gy) | $D_{max}/D_{Rx}$(%) |
| Nominal | 94.80 | 63.06 | 68.23 | 96.25 | 61.26 | 66.10 | 99.12 | 54.69 | 55.38 | 2.36 | 18.86 | 6.55 | 111.96 |
| Online | 93.29 | 64.29 | 67.08 | 96.50 | 62.84 | 66.38 | 99.32 | 55.05 | 55.88 | 2.07 | 18.49 | 6.44 | 109.68 |
| Offline | 92.07 | 62.20 | 65.91 | 94.72 | 60.78 | 64.62 | 97.88 | 53.93 | 54.82 | 1.64 | 19.60 | 6.53 | 113.75 |
| Patient 2 | CTV-High | | | CTV-Mid | | | CTV-Low | | | Right Parotid | Left Parotid | Oral Cavity | Body |
| | $V_{100}$(%) | $D_{98}$(Gy) | $D_{95}$(Gy) | $V_{100}$(%) | $D_{98}$(Gy) | $D_{95}$(Gy) | $V_{100}$(%) | $D_{98}$(Gy) | $D_{95}$(Gy) | $D_{mean}$(Gy) | $D_{mean}$(Gy) | $D_{mean}$(Gy) | $D_{max}/D_{Rx}$(%) |
| Nominal | 98.20 | 70.14 | 70.98 | 99.17 | 64.63 | 66.92 | 99.08 | 56.66 | 57.30 | 18.80 | 24.62 | 30.30 | 106.46 |
| Online | 98.88 | 70.62 | 71.44 | 99.81 | 66.85 | 68.34 | 98.97 | 56.72 | 57.41 | 15.66 | 23.50 | 29.34 | 104.24 |
| Offline | 95.20 | 69.02 | 70.05 | 98.64 | 63.62 | 65.94 | 89.53 | 51.88 | 54.47 | 17.96 | 24.46 | 30.63 | 106.29 |
| Patient 3 | CTV-High | | | CTV-Mid | | | CTV-Low | | | Right Parotid | Left Parotid | Oral Cavity | Body |
| | $V_{100}$(%) | $D_{98}$(Gy) | $D_{95}$(Gy) | $V_{100}$(%) | $D_{98}$(Gy) | $D_{95}$(Gy) | $V_{100}$(%) | $D_{98}$(Gy) | $D_{95}$(Gy) | $D_{mean}$(Gy) | $D_{mean}$(Gy) | $D_{mean}$(Gy) | $D_{max}/D_{Rx}$(%) |
| Nominal | 99.89 | 71.06 | 71.42 | 98.47 | 60.40 | 61.11 | 99.08 | 54.38 | 54.92 | 28.58 | 19.17 | 37.12 | 107.36 |
| Online | 98.82 | 70.46 | 71.01 | 97.62 | 59.94 | 61.12 | 98.99 | 54.40 | 54.82 | 22.16 | 15.87 | 34.61 | 104.41 |
| Offline | 98.17 | 70.05 | 70.51 | 93.83 | 58.78 | 59.92 | 99.10 | 54.42 | 54.93 | 29.52 | 20.12 | 41.89 | 109.19 |
| Patient 4 | CTV-High | | | CTV-Mid | | | CTV-Low | | | Right Parotid | Left Parotid | Oral Cavity | Body |
| | $V_{100}$(%) | $D_{98}$(Gy) | $D_{95}$(Gy) | $V_{100}$(%) | $D_{98}$(Gy) | $D_{95}$(Gy) | $V_{100}$(%) | $D_{98}$(Gy) | $D_{95}$(Gy) | $D_{mean}$(Gy) | $D_{mean}$(Gy) | $D_{mean}$(Gy) | $D_{max}/D_{Rx}$(%) |
| Nominal | 98.23 | 70.17 | 70.82 | 98.80 | 63.32 | 63.74 | 98.44 | 56.20 | 56.56 | 21.19 | 19.81 | 2.83 | 106.40 |
| Online | 95.66 | 69.27 | 70.18 | 97.73 | 62.92 | 63.42 | 96.09 | 55.21 | 56.21 | 19.37 | 17.15 | 2.85 | 103.70 |
| Offline | 97.24 | 69.76 | 70.47 | 98.84 | 63.60 | 64.41 | 96.35 | 55.35 | 56.26 | 21.71 | 20.25 | 5.72 | 105.90 |

### 3.2. TCP and NTCP

For each patient, we have calculated TCP and NTCP for the nominal plan, the simulated online APT course, and the actual offline APT course. The discrepancies of the calculated TCP and NTCP values between the nominal plan and the two APT courses (denoted as ΔTCP and ΔNTCP) are reported in this manuscript to better illustrate the potential impact of the online APT and the offline APT on treatment outcome. Please note that the NTCP calculated for both APT courses were comparable to that of the nominal plan for patients 1 and 2, and the TCP for both APT courses were comparable to that of the nominal plan for patients 3 and 4. Hence, Figure 6 depicts ΔTCP for the two APT courses only for patients 1 and 2, and Figure 7 depicts ΔNTCP for patients 3 and 4. It can be observed from these two figures that patients 1 and 2 might benefit from online APT in terms of improved local tumor control, particularly for CTV-Low with an improvement of 2.66% and 4.52% in TCP, respectively. Patient 3 might benefit from online APT in terms of reduced treatment toxicity with a derease of 1.67% and 3.40% in NCTP for right parotid and oral cavity. Patient 4 might benefit from online APT in terms of reduced treatment toxicity with a derease of 0.73% in NCTP for right parotid.

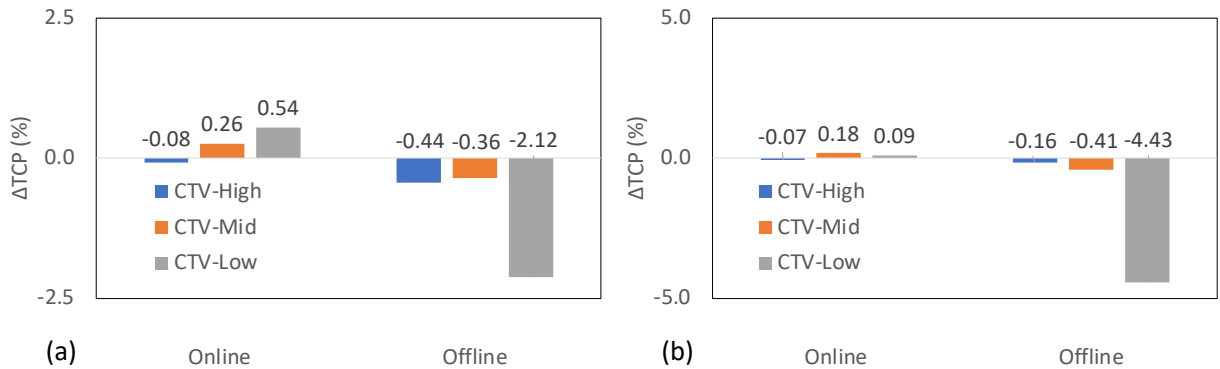

**Figure 6.** ΔTCP between the nominal plan and the two APT courses for patients 1 and 2.

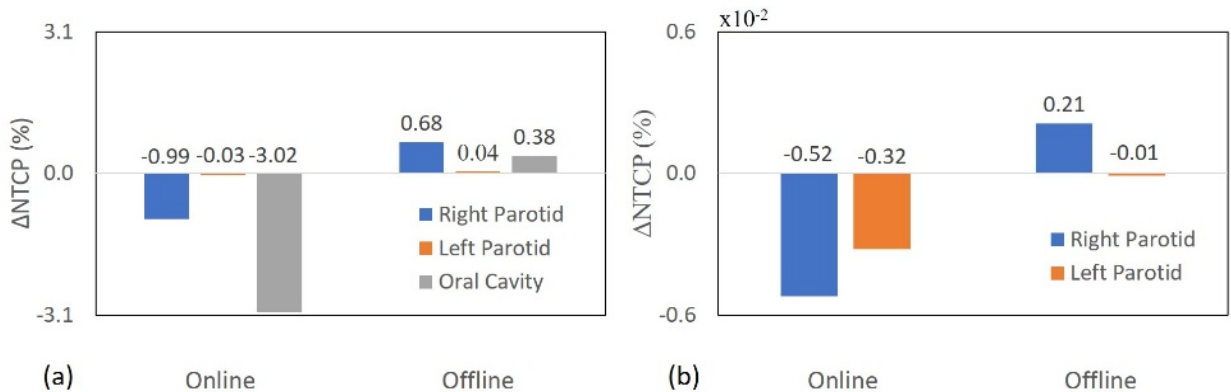

**Figure 7.** ΔNTCP between the nominal plan and the two APT courses for patients 3 and 4.

### 4. Discussions

The CBCT images acquired by the existing on-board imaging system have poor soft tissue contrast and large HU uncertainties, and hence cannot be directly used for daily dose evaluation and online plan adaptation. In our retrospective study, we deformed the planning CT images to the daily CBCT images and copied the air cavities from the CBCT images to the deformed planning CT images, in order to use the correct HU values from the planning CT images while persevering the actual patient anatomy captured on the CBCT images. This strategy is subject to the accuracy of the image deformable registration. For this particular study, our purpose is to compare the efficacy of the online APT and the offline APT with the presence of substantial anatomical changes. Even if the corrected CBCT images didn't exactly represent the actual patient anatomy due to the imperfect deformable registration, as the same images were used to accumulate the actual dose for both online and offline APT courses, our comparisons were fair enough to yield reasonable conclusions. Efforts have been devoted to improving the CBCT image quality and HU accuracy by generating synthetic CT images from the CBCT images using advanced deep learning technologies [55-60]. This may be an alternative solution for CBCT-guided online APT to eliminate the impact of the deformable registration accuracy on the daily dose evaluation and online plan adaptation.

As manual contouring is very laborious and not feasible for online APT, in this study we chose to propagate the OAR contours from the planning CT images onto the corrected CBCT images by rigidly copying or deformable mapping contours of different structures, which is similar to the contour propagation from the planning CT images onto the QACT images in the current clinical practice of offline APT. With deep learning methods of CBCT-based synthetic imaging [61, 62], automatic organ segmentation might be an alternative way to achieve automatic OAR contouring for online APT [63-65]. As the logistics and cost prevent acquiring daily MRI during the 4-6 week course of treatment to examine the tumor regression status and obtain the actual CTV volumes, the original CTV contours that were manually delineated by physicians on the planning CT images were also propagated onto either the corrected CBCT images for online APT or the QACT image for offline APT via image deformable registration [66], assuming that features and relations between the CTVs and the nearby anatomy structures that were present on the planning CT images were preserved on the images of the patient's new anatomy. The same assumption was used by Ethos for CTV contour propagation in CBCT-based online adaptive photon therapy [67]. This CTV contour propagation strategy may not be able to seamlessly incorporate the tumor volume change into plan adaptation for patients with rapid tumor volume change. For these patients, MRI images may need to be acquired more frequently (e.g., weekly) along the treatment course to help the CTV delineation on the corrected CBCT images for both online and offline adaptation. Similarly, as the same CTV propagation strategy was used for both online and offline adaptation and the same CTV contours that have been reviewed by physicians were used for dose accumulation in both courses, our comparisons between the online and offline APT courses in this particular study were fair enough to yield reasonable conclusions.

In our retrospective study, due to the unavailability of the online APT technology, the replanning process was actually performed by our proton dosimetrist but was regarded to be a "online" process to mimic the online APT, by assuming that automatic and rapid online replanning will be realized via advanced artificial intelligence technologies in future to achieve sufficient plan quality that is comparable to the plans generated by human planners. This assumption was made based on the recent success of deep reinforcement learning based automatic treatment planning for IMRT for prostate cancer [68-70] and pancreas cancer [71], brachytherapy for cervical cancer [72, 73], and Gamma Knife radiosurgery for vestibular schwannoma [74]. It is important to note that the observed improvements achieved by the online APT course are a combination of online adaptation based on patient's daily treatment anatomy, and the strategy of including only three treatment scenarios in the robust optimization to account for the range uncertainty and using a small margin (e.g., 1.5 mm) to account for potential intra-fractional motion and spot positioning uncertainty instead of using 21 treatment scenarios generated with a same range uncertainty but a large setup uncertainty (e.g., 3 mm). This strategy can only be safely used in the context of online adaptation because of the much-increased certainty in target and OAR positioning at each treatment fraction, as the actual patient anatomy in treatment position for the current fraction has been captured on the CBCT images.

The results of our retrospective study have illustrated that some patients may benefit from online adaptation more than others and may be in different aspects, depending on the magnitude of the patient's anatomical changes and the proximity of the critical organs. For instance, patients with significant and rapid anatomical changes are expected to benefit more from online adaptation than patients with moderate and relatively slow anatomical changes. Besides, for patients who have one or multiple critical and radiosensitive organs very close to CTVs, it may be very challenging to achieve a high-quality nominal plan and the planned dose of those critical organs might already be on the borderline. Even a small anatomical change might degrade the actual plan quality of the nominal plan, leading to the failing of the clinical goals for those critical organs. Hence, these patients with a challenging anatomy for treatment planning are also expected to benefit more from online adaptation than patients who have a much easier anatomy that results in a very high-quality nominal plan passing the clinical goals by a large margin. Because of the extra clinical resources (e.g., time and personnel) required for online APT, it would be ideal to identify the patients with large potential benefits prior to treatment and triage them for online APT, while other patients continue with current non-adaptive or offline adaptive proton therapy. One limitation of our retrospective study is the small amount of patient cases, due to the current heavy workload involved in this study without any available online APT systems, such as deformable registration of the planning CT images to the daily CBCT images and contour propagation, review, and modification at every treatment fraction, dose recalculation for the in-use plan of the online APT course and the in-use plan of the offline APT course at very fraction, dose evaluation at every fraction of the online APT course, replanning whenever online adaptation is triggered, deformable registration of the CBCT images of every fraction to the planning CT images [75], and dose accumulation over every fraction for both online and offline APT courses. In our future work, we will collect more patient cases and continue this retrospective study. With the results of enough patient cases, we will examine the characteristics of those patients who show big benefit from online adaptation to establish whether or not we can identify the patients who are mostly likely to benefit from online adaptation.

## 5. Conclusions

To investigate the potential effect of online APT and compare it with the current offline APT practice, we have performed a retrospective study on four HN cancer patients, who were treated with proton therapy at our institution and had at least one plan adaptation during their actual treatment courses, simulating a CBCT-guide online APT course using the daily CBCT images acquired from each treatment fraction of the actual offline APT course. Among the four patient cases, three cases were found to benefit from online APT versus offline APT to different extents. For the fourth patient case, as both courses show their improvement in certain planning objectives and deterioration in other objectives when compared to the other, which might be due to the different priorities of the planning objectives used in the two courses, it is difficult to determine whether this patient can benefit from online APT or not. Notably, online APT tends to spare more healthy tissues by achieving the clinical goal with merely half of the positional uncertainty margin, compared to the nominal clinical plan. As some patients can benefit more from online APT than others, depending on the magnitude of the patient's anatomical changes and the proximity of the critical organs to the treatment targets, we think it is still worth devoting great efforts to developing advanced algorithms to enable online APT. While, future studies are needed to help identify the patients with large potential benefits from online APT prior to treatment in order to conserve scarce clinical resources.

## Acknowledgments

This research is partly supported by the National Institutes of Health under Award Number R01CA215718 (XY) and R37CA272755 (ZT).

## References

1. Siegel, R.L., et al., *Cancer statistics, 2021.* Ca Cancer J Clin, 2021. **71**(1): p. 7-33.
2. van Vulpen, M., et al., *Comparing step-and-shoot IMRT with dynamic helical tomotherapy IMRT plans for head-and-neck cancer.* International Journal of Radiation Oncology* Biology* Physics, 2005. **62**(5): p. 1535-1539.
3. Mendenhall, W.M., R.J. Amdur, and J.R. Palta, *Intensity-modulated radiotherapy in the standard management of head and neck cancer: promises and pitfalls.* Journal of clinical oncology, 2006. **24**(17): p. 2618-2623.
4. Verbakel, W.F., et al., *Volumetric intensity-modulated arc therapy vs. conventional IMRT in head-and-neck cancer: a comparative planning and dosimetric study.* International Journal of Radiation Oncology* Biology* Physics, 2009. **74**(1): p. 252-259.
5. Gomez-Millan, J., J.R. Fernández, and J.A.M. Carmona, *Current status of IMRT in head and neck cancer.* Reports of Practical Oncology & Radiotherapy, 2013. **18**(6): p. 371-375.
6. Gomez–Millan Barrachina, J., et al., *Potential advantages of volumetric arc therapy in head and neck cancer.* Head & neck, 2015. **37**(6): p. 909-914.
7. Simone II, C.B., et al., *Comparison of intensity-modulated radiotherapy, adaptive radiotherapy, proton radiotherapy, and adaptive proton radiotherapy for treatment of locally advanced head and neck cancer.* Radiotherapy and Oncology, 2011. **101**(3): p. 376-382.
8. Moreno, A.C., et al., *Intensity modulated proton therapy (IMPT) - The future of IMRT for head and neck cancer.* Oral Oncol, 2019. **88**: p. 66-74.
9. Nguyen, M.L., et al., *Intensity-modulated proton therapy (IMPT) versus intensity-modulated radiation therapy (IMRT) for the treatment of head and neck cancer: A dosimetric comparison.* Med Dosim, 2021. **46**(3): p. 259-263.
10. Holliday, E.B., et al. *Proton Therapy Reduces Treatment-Related Toxicities for Patients with Nasopharyngeal Cancer: A Case-Match Control Study of Intensity-Modulated Proton Therapy and Intensity-Modulated Photon Therapy.* 2015.
11. Blanchard, P., et al., *Intensity-modulated proton beam therapy (IMPT) versus intensity-modulated photon therapy (IMRT) for patients with oropharynx cancer - A case matched analysis.* Radiother Oncol, 2016. **120**(1): p. 48-55.
12. McKeever, M.R., et al., *Reduced acute toxicity and improved efficacy from intensity-modulated proton therapy (IMPT) for the management of head and neck cancer.* Chin Clin Oncol, 2016. **5**(4): p. 54.
13. Sharma, S., et al., *Quality of Life of Postoperative Photon versus Proton Radiation Therapy for Oropharynx Cancer.* Int J Part Ther, 2018. **5**(2): p. 11-17.
14. Li, X., et al., *Toxicity Profiles and Survival Outcomes Among Patients With Nonmetastatic Nasopharyngeal Carcinoma Treated With Intensity-Modulated Proton Therapy vs Intensity-Modulated Radiation Therapy.* JAMA Netw Open, 2021. **4**(6): p. e2113205.
15. Smith, G.L., et al., *Work Outcomes after Intensity-Modulated Proton Therapy (IMPT) versus Intensity-Modulated Photon Therapy (IMRT) for Oropharyngeal Cancer.* Int J Part Ther, 2021. **8**(1): p. 319-327.
16. Barker Jr, J.L., et al., *Quantification of volumetric and geometric changes occurring during fractionated radiotherapy for head-and-neck cancer using an integrated CT/linear accelerator system.* International Journal of Radiation Oncology* Biology* Physics, 2004. **59**(4): p. 960-970.
17. Paganetti, H., *Range uncertainties in proton therapy and the role of Monte Carlo simulations.* Physics in Medicine and Biology, 2012. **57**(11): p. R99-R117.
18. Chang, C.-W., et al., *A standardized commissioning framework of Monte Carlo dose calculation algorithms for proton pencil beam scanning treatment planning systems.* Medical Physics, 2020. **47**(4): p. 1545-1557.
19. Chang, C.-W., et al., *Dual-energy CT based mass density and relative stopping power estimation for proton therapy using physics-informed deep learning.* Physics in Medicine & Biology, 2022. **67**(11): p. 115010.
20. Chang, C.-W., et al., *Validation of a deep learning-based material estimation model for Monte Carlo dose calculation in proton therapy.* Physics in Medicine & Biology, 2022. **67**(21): p. 215004.

21. Liu, W., et al., *Robust optimization of intensity modulated proton therapy.* Medical Physics, 2012. **39**(2): p. 1079-1091.
22. Li, Y., et al., *Selective robust optimization: a new intensity-modulated proton therapy optimization strategy.* Medical physics, 2015. **42**(8): p. 4840-4847.
23. Liu, W., et al., *Effectiveness of robust optimization in intensity-modulated proton therapy planning for head and neck cancers.* Medical physics, 2013. **40**(5): p. 051711.
24. van Dijk, L.V., et al., *Robust intensity modulated proton therapy (IMPT) increases estimated clinical benefit in head and neck cancer patients.* PLoS One, 2016. **11**(3): p. e0152477.
25. Zhou, J., et al., *Dosimetric Uncertainties in Dominant Intraprostatic Lesion Simultaneous Boost Using Intensity Modulated Proton Therapy.* Advances in Radiation Oncology, 2022. **7**(1): p. 100826.
26. Cubillos-Mesías, M., et al., *Impact of robust treatment planning on single-and multi-field optimized plans for proton beam therapy of unilateral head and neck target volumes.* Radiation Oncology, 2017. **12**(1): p. 1-10.
27. Cubillos-Mesías, M., et al., *Including anatomical variations in robust optimization for head and neck proton therapy can reduce the need of adaptation.* Radiotherapy and Oncology, 2019. **131**: p. 127-134.
28. Lalonde, A., et al., *Anatomic changes in head and neck intensity-modulated proton therapy: Comparison between robust optimization and online adaptation.* Radiother Oncol, 2021. **159**: p. 39-47.
29. Buti, G., et al., *Accelerated robust optimization algorithm for proton therapy treatment planning.* Medical Physics, 2020. **47**(7): p. 2746-2754.
30. Botas, P., et al., *Online adaption approaches for intensity modulated proton therapy for head and neck patients based on cone beam CTs and Monte Carlo simulations.* Physics in Medicine & Biology, 2018. **64**(1): p. 015004.
31. Lalonde, A., et al., *Anatomic changes in head and neck intensity-modulated proton therapy: Comparison between robust optimization and online adaptation.* Radiotherapy and Oncology, 2021. **159**: p. 39-47.
32. Chang, C.-W., et al., *Early in vivo Radiation Damage Quantification for Pediatric Craniospinal Irradiation Using Longitudinal MRI for Intensity Modulated Proton Therapy.* arXiv preprint arXiv:2210.15557, 2022.
33. Stanforth, A., et al., *Onboard cone-beam CT-based replan evaluation for head and neck proton therapy.* Journal of Applied Clinical Medical Physics, 2022. **23**(5): p. e13550.
34. Chang, C.-W., et al., *Deep learning-based Fast Volumetric Image Generation for Image-guided Proton FLASH Radiotherapy.* arXiv preprint arXiv:2210.00971, 2022.
35. Troost, E., et al., *Towards online adaptive proton therapy: first report of plan-library-based plan-of-the-day approach.* Acta Oncologica, 2022. **61**(2): p. 231-234.
36. Paganetti, H., et al., *Adaptive proton therapy.* Physics in Medicine & Biology, 2021. **66**(22): p. 22TR01.
37. Elmahdy, M.S., et al., *Robust contour propagation using deep learning and image registration for online adaptive proton therapy of prostate cancer.* Medical physics, 2019. **46**(8): p. 3329-3343.
38. Albertini, F., et al., *Online daily adaptive proton therapy.* The British journal of radiology, 2020. **93**(1107): p. 20190594.
39. Villarroel, E.B., X. Geets, and E. Sterpin, *Online adaptive dose restoration in intensity modulated proton therapy of lung cancer to account for inter-fractional density changes.* Physics and imaging in radiation oncology, 2020. **15**: p. 30-37.
40. Nenoff, L., et al., *Integrating Structure Propagation Uncertainties in the Optimization of Online Adaptive Proton Therapy Plans.* Cancers, 2022. **14**(16): p. 3926.
41. Nesteruk, K.P., et al., *CT-On-Rails Versus In-Room CBCT for Online Daily Adaptive Proton Therapy of Head-And-Neck Cancers.* Cancers, 2021. **13**(23): p. 5991.
42. Feng, H., et al., *GPU-accelerated Monte Carlo-based online adaptive proton therapy: A feasibility study.* Medical Physics, 2022. **49**(6): p. 3550-3563.
43. ICRU78, *Prescribing, Recording, and Reporting Proton-Beam Therapy.* ICRU Publication 78, 2007.
44. *Relative Biological Effectiveness in Ion Beam Therapy.* 2008, Vienna: INTERNATIONAL ATOMIC ENERGY AGENCY.

45. Kang, C.L., et al., *Comparison of Intrafractional Motion in Head and Neck Cancer Between Two Immobilization Methods During Stereotactic Ablative Radiation Therapy by CyberKnife.* Cancer Manag Res, 2020. **12**: p. 13599-13606.
46. Li, H., et al., *SU-FF-T-288: Spot Position Uncertainty of Scanning Proton Pencil Beam.* Medical physics, 2009. **36**(6Part13): p. 2587-2587.
47. Gay, H.A. and A. Niemierko, *A free program for calculating EUD-based NTCP and TCP in external beam radiotherapy.* Physica Medica, 2007. **23**(3): p. 115-125.
48. Williams, M.V., J. Denekamp, and J.F. Fowler, *A review of αβ ratios for experimental tumors: Implications for clinical studies of altered fractionation.* International Journal of Radiation Oncology*Biology*Physics, 1985. **11**(1): p. 87-96.
49. Burman, C., et al., *Fitting of normal tissue tolerance data to an analytic function.* International Journal of Radiation Oncology*Biology*Physics, 1991. **21**(1): p. 123-135.
50. Emami, B., et al., *Tolerance of normal tissue to therapeutic irradiation.* International Journal of Radiation Oncology*Biology*Physics, 1991. **21**(1): p. 109-122.
51. Budach, W., et al., *The TCD50 and regrowth delay assay in human tumor xenografts: Differences and implications.* International Journal of Radiation Oncology*Biology*Physics, 1993. **25**(2): p. 259-268.
52. Wu, Q., et al., *Optimization of intensity-modulated radiotherapy plans based on the equivalent uniform dose.* International Journal of Radiation Oncology, Biology, Physics, 2002. **52**(1): p. 224-235.
53. van Leeuwen, C.M., et al., *The alfa and beta of tumours: a review of parameters of the linear-quadratic model, derived from clinical radiotherapy studies.* Radiation Oncology, 2018. **13**(1): p. 96.
54. Hamming-Vrieze, O., et al., *Impact of setup and range uncertainties on TCP and NTCP following VMAT or IMPT of oropharyngeal cancer patients.* Physics in Medicine & Biology, 2019. **64**(9): p. 095001.
55. Harms, J., et al., *Paired cycle-GAN-based image correction for quantitative cone-beam computed tomography.* Medical Physics, 2019. **46**(9): p. 3998-4009.
56. Lei, Y., et al., *Learning-based CBCT correction using alternating random forest based on auto-context model.* Medical Physics, 2019. **46**(2): p. 601-618.
57. Lalonde, A., et al., *Evaluation of CBCT scatter correction using deep convolutional neural networks for head and neck adaptive proton therapy.* Physics in Medicine & Biology, 2020. **65**(24): p. 245022.
58. Liu, Y., et al., *CBCT-based synthetic CT generation using deep-attention cycleGAN for pancreatic adaptive radiotherapy.* Medical physics, 2020. **47**(6): p. 2472-2483.
59. Thummerer, A., et al., *Clinical suitability of deep learning based synthetic CTs for adaptive proton therapy of lung cancer.* Medical physics, 2021. **48**(12): p. 7673-7684.
60. Uh, J., et al., *Training a deep neural network coping with diversities in abdominal and pelvic images of children and young adults for CBCT-based adaptive proton therapy.* Radiotherapy and Oncology, 2021. **160**: p. 250-258.
61. Lei, Y., et al. *CBCT-Based Synthetic MRI Generation for CBCT-Guided Adaptive Radiotherapy*. in *Artificial Intelligence in Radiation Therapy*. 2019. Cham: Springer International Publishing.
62. Wang, T., et al., *A review on medical imaging synthesis using deep learning and its clinical applications.* Journal of Applied Clinical Medical Physics, 2021. **22**(1): p. 11-36.
63. Dai, X., et al., *Synthetic MRI-aided Head-and-Neck Organs-at-Risk Auto-Delineation for CBCT-guided Adaptive Radiotherapy.* arXiv preprint arXiv:2010.04275, 2020.
64. Dai, X., et al., *Automated delineation of head and neck organs at risk using synthetic MRI-aided mask scoring regional convolutional neural network.* Medical Physics, 2021. **48**(10): p. 5862-5873.
65. Dai, X., et al., *Synthetic CT-aided multiorgan segmentation for CBCT-guided adaptive pancreatic radiotherapy.* Medical Physics, 2021. **48**(11): p. 7063-7073.
66. Lei, Y., et al., *Deformable CT image registration via a dual feasible neural network.* Medical Physics, 2022. **n/a**(n/a).
67. Archambault, Y., et al., *Making on-line adaptive radiotherapy possible using artificial intelligence and machine learning for efficient daily re-planning.* Med Phys Intl J, 2020. **8**(2).
68. Shen, C., et al., *Operating a treatment planning system using a deep-reinforcement learning-based virtual treatment planner for prostate cancer intensity-modulated radiation therapy treatment planning.* Medical physics, 2020. **47**(6): p. 2329-2336.

69. Shen, C., L. Chen, and X. Jia, *A hierarchical deep reinforcement learning framework for intelligent automatic treatment planning of prostate cancer intensity modulated radiation therapy.* Physics in Medicine & Biology, 2021. **66**(13): p. 134002.
70. Sprouts, D., et al., *The development of a deep reinforcement learning network for dose-volume-constrained treatment planning in prostate cancer intensity modulated radiotherapy.* Biomedical Physics & Engineering Express, 2022. **8**(4): p. 045008.
71. Zhang, J., et al., *An interpretable planning bot for pancreas stereotactic body radiation therapy.* International Journal of Radiation Oncology* Biology* Physics, 2021. **109**(4): p. 1076-1085.
72. Shen, C., et al., *Intelligent inverse treatment planning via deep reinforcement learning, a proof-of-principle study in high dose-rate brachytherapy for cervical cancer.* Physics in Medicine & Biology, 2019. **64**(11): p. 115013.
73. Pu, G., et al., *Deep reinforcement learning for treatment planning in high-dose-rate cervical brachytherapy.* Physica Medica, 2022. **94**: p. 1-7.
74. Liu, Y., et al., *Automatic inverse treatment planning of Gamma Knife radiosurgery via deep reinforcement learning.* Medical Physics, 2022. **49**(5): p. 2877-2889.
75. Xie, H., et al., *Deformable Image Registration using Unsupervised Deep Learning for CBCT-guided Abdominal Radiotherapy.* arXiv preprint arXiv:2208.13686, 2022.